\documentclass[aps,pra,twocolumn,showpacs,groupedaddress]{revtex4}
\usepackage{graphicx}

\begin{document}

%\draft

\title{A fundamental limit for integrated atom optics with Bose-Einstein condensates}
\author{Weiping Zhang, Ewan M. Wright, Han Pu, and Pierre Meystre}
\affiliation{Optical Sciences Center, University of Arizona,
Tucson, AZ 85721}

\date{\today}

\begin{abstract}
The dynamical response of an atomic Bose-Einstein condensate
manipulated by an integrated atom optics device such as a
microtrap or a microfabricated waveguide is studied. We show that
when the miniaturization of the device enforces a sufficiently
high condensate density, three-body interactions lead to a
spatial modulational instability that results in a fundamental
limit on the coherent manipulation of Bose-Einstein condensates.
\end{abstract}

 \maketitle

\section{Introduction}
The central idea of integrated atom optics is to miniaturize the
size of active and passive atom optical components such as atom
lasers, wave guides, beam splitters, and interferometers, and to
combine them in integrated devices
\cite{Weinstein,Fortagh,Reichel,Prentiss,Muller,Dekker,Hinds,
Folman,Andersson,Fortagha,ketterle,Jones}. Potential future
applications include hand-carried high-precision measurement
devices such as gravimeters and gyroscopes. Integrated
atom-optical chips may also prove useful to control neutral atoms
in integrated and scalable microtrap arrays for quantum
information processing.  For example, microfabricated atom
optical components can be fabricated with current-carrying wires
or microstructured surfaces, combined with homogeneous magnetic
bias fields and periodically magnetized substrates. Optical
fields or electric fields may also be involved in the hybrid
construction of such devices.

However, the future of integrated atom optics still depends on
finding solutions of a number of both technical and fundamental
issues. For example, it will be essential to maintain the
coherence of the matter waves to be manipulated, a step still to
be demonstrated in waveguide-based beam splitters. Atoms confined
in wire-based microtraps on atom chips face a relatively noisy
environment, due e.g. to the heating from the chip substrate,
which can cause the decoherence of atomic matter wave
\cite{Henkel}. In addition, fluctuating currents in micro-wires
may perturb the atoms when they are brought near its surface:
Recent experiments have revealed the fragmentation of atomic
Bose-Einstein condensates (BECs) confined in wire-based
microtraps when the atom-surface distance is reduced to
micrometer scale \cite{Fortagha,ketterle,Jones}, and the
experimental data seems to support the argument that the
fluctuating currents in the wire and corrugations due to
imperfect microfabrication are responsible for that fragmentation.

While such experiments illustrate the technical limitations for
the coherent control of atoms in the microtraps, it can be hoped
that these will be eliminated by better engineering solutions. In
addition, though, it is also important to understand the
fundamental limitations to integrated atom optics. For example,
the strong compression typical of microtraps or waveguides can
significantly increase the density of the atomic BECs, thereby
enhancing inelastic collisions and resulting in a fundamental
source of decoherence that reduces the lifetime of the confined
BECs \cite{Fortagha}.

In this paper, we discuss an additional effect that limits the
coherent control of atomic BECs in microtraps or microfabricated
waveguides. Specifically, we show that when the miniaturization of
the device results in a sufficiently high condensate density,
three-body interactions lead to a spatial modulational instability
that leads to the fragmentation of the condensate, thereby
imposing a fundamental limit on the densities that can be stably
propagated and manipulated in atomic waveguides.

The paper is organized as follows: Section II introduces our model
and establishes the notation. Section III uses a linear stability
analysis to show analytically the existence of a modulational
instability in the condensate dynamics when three-body collisions
become important. Section IV presents the results of selected
numerical simulations that confirm the analytical predictions and
show the onset of condensate fragmentation. Section V expands our
analysis and presents arguments that reinforce the conclusions of
the numerical simulations and argue that the modulational
instability eventually leads to a collapse of the condensate.
Finally, section VI is a summary and outlook.
\section{The model}
For concreteness, we consider in this paper an atomic waveguide
resulting from the combination of a current-carrying wire (for
instance) electroplated on a substrate, and a static, homogeneous
bias magnetic field $B_{\rm bias}$ \cite{Fortagha} perpendicular
to the conductor. Assuming a linear wire, the distance $d$ of the
waveguide to the chip surface is
\begin{eqnarray}
d=\left (\frac{\mu_0}{2 \pi}\right )  \frac{I}{B_{\rm bias}} ,
\end{eqnarray}
where $\mu_0$ is the vacuum permeability and $I$ the current
through the wire. Since the guide height can be reduced using
smaller currents or stronger bias field, the setup is ideal for
the requirement of miniaturization in integrated atom optics.

The radial oscillation frequency of this trap is
\begin{equation}
\omega_r = \frac{2\pi}{\mu_0} \sqrt{\frac{\mu_B g_F
m_F}{mB_0}}\frac{B_{\rm bias}^2}{I},
\end{equation}
where $\mu_B$ is the Bohr magneton, $g_F$ and $m_F$ the
gyromagnetic ratio and magnetic moment of the hyperfine state of
the alkali atoms of mass $m$ being trapped, and we have assumed
that the axial confinement of the atoms is achieved by a
Ioffe-Pritchard-type trap that provides a field $B_0$ in the
center of the trap.

Since the radial oscillation frequency $\omega_r$ scales as
$B_{\rm bias}^2/I$, any miniaturization of the trap achieved
through a reduction of $d$ results in an increase of $\omega_r$,
and hence in the density of the confined condensate. The closer
to the chip surface the condensate is brought, the higher its
density. In some of the current experiments, the condensate
densities can already reach values in excess of $1 \times
10^{15}$ cm$^{-3}$. For such high densities, the effects of
three-body interactions on the dynamics of BECs are expected to
become important. Our goal in this paper is to understand their
impact on the dynamics of the guided condensate.

Our starting point is the three-dimensional Gross-Pitaevskii
equation for the wave function $\Psi({\bf r},t)$ of a trapped
condensate, generalized to include three-body interactions,
\begin{eqnarray}
i\hbar \frac{\partial \Psi}{\partial t} &=&
\left[-\frac{\hbar^2}{2m}\nabla^2  +
\frac{1}{2} m \omega_z^2[\xi^2(t)r^2 + z^2] -\mu \right]\Psi \nonumber \\
&& + \hbar g_2 N|\Psi|^2 \Psi +\hbar g_3 N^2 |\Psi|^4 \Psi.
\end{eqnarray}
Here $\mu$ is the chemical potential, $g_2$ and $g_3$ measure the
strength of the two-body and three-body interactions,
respectively, $\Psi({\bf r},t)$ is normalized to unity and $N$ is
the number of atoms in the condensate. We have also introduced
the dimensionless ratio
\begin{equation}
\xi(t)=\omega_r(t)/\omega_z,
\end{equation}
which determines the strength of the transverse confinement
relative to the longitudinal trapping. (We allow $\xi(t)$ to be
time-dependent to include possible time variations of bias
magnetic field $B_{\rm bias}$ or of the current $I$, and hence of
$\omega_r$.)

In the following, we assume that the microtrap is operating in
the regime of tight transverse confinement, $\xi(t)>> 1$ and also
that the transverse trapping potential is much larger than the
interatomic interaction. We further assume that the time
variation of $\omega_r(t)$ is slow enough that the transverse
profile of the BEC adiabatically follows the ground transverse
oscillator state
\begin{equation}
\phi_0(r,t)=\sqrt{{1 \over \pi l_r^2(t)}}e^{-r^2/2l_r^2(t)} ,
\end{equation}
with $l_r(t)=\sqrt{\hbar/m\omega_r(t)}$.

We can then proceed by approximating the condensate wave function
$\Psi({\bf r},t)$ as
\begin{equation}
\Psi({\bf r},t) \approx u(z,t) \phi_0(r,t)e^{-i\int^t
dt'\omega_r(t')} ,
\end{equation}
where it is understood that the radial wave function
$\phi_0(r,t)$ contains only slow time variations, and the
envelope function $u(z,t)$ incorporates all other time
variations. We then obtain the approximate equation for the
envelope function
\begin{eqnarray}
i \hbar \frac{\partial u}{\partial t} &=&
\left[-\frac{\hbar^2}{2m} \frac{\partial^2}{\partial z^2}  +
\frac{1}{2} m \omega_z^2 z^2 -\mu \right]u \nonumber \\
&& + \hbar g_{2,z}(t)|u|^2 u+\hbar g_{3,z}(t) |u|^4 u,
\label{1dgpeq}
\end{eqnarray}
where the reduced interatomic interaction coefficients $g_{2,z}$
and $g_{3,z}$ are
\begin{eqnarray}
g_{2,z}(t) &=& g_2 N \int 2\pi rdr|\phi_0(r,t)|^4 = \left
(\frac{g_2 N}{2\pi l_z^2}\right ) \xi(t), \nonumber
\\ g_{3,z}(t) &=& g_3 N^2 \int 2\pi rdr|\phi_0(r,t)|^6 = \left
(\frac{g_3N^2}{3\pi^2 l_z^4}\right ) \xi^2(t) ,
\end{eqnarray}
with $l_z=\sqrt{\hbar/m\omega_z}$. From these definitions we see
that the reduced two-body interaction coefficient varies linearly
with the transverse oscillator frequency or the ratio
$\xi(t)=\omega_t(t)/\omega_z$, whereas the three-body coefficient
scales as $\xi^2(t)$. This implies that the relative significance
of two- and three-body interactions depends on the level of
transverse confinement.

For alkali atoms, the two-body interaction coefficient has fully
been determined by experiments. Although information for
three-body interactions is still sketchy, recent theoretical
studies have provided a way to estimate the three-body
interaction coefficient $g_3$
\cite{Bedaque,Braatena,Braatenb,Jack,Kohler}. For rubidium atoms,
it may be expected to be negative, with $|g_3|$ of the order of
$10^{-26} - 10^{-27}$ cm$^6$/s. With this in mind, we focus our
discussion on atoms with repulsive two-body interactions ($g_2
> 0$) and attractive three-body interactions ($g_3 < 0$).
\section{Modulational instability}
To expose the basic modulational instability that arises at high
densities we first consider the case where the longitudinal
trapping potential is switched off, the transverse trapping is
fixed, so that $\xi>>1$ is constant, and the condensate has a
homogeneous longitudinal distribution $u_0$ over a length $L$. The
chemical potential is then $\mu=\hbar g_{2,z} |u_0|^2 -\hbar
|g_{3,z}| |u_0|^4$.

We are interested in the condensate dynamics in the presence of
the attractive three-body interaction. For this purpose, we first
carry out a linear stability analysis by setting
\begin{equation}
u(z,t)=[u_0 + \delta u(z,t)]e^{-i\mu t/\hbar}, \label{excitationa}
\end{equation}
and looking at longitudinal excitations of the form
\begin{eqnarray*}
\delta u(z,t)= \sum_k \left[\alpha_k U_k(z) e^{-i\nu_k
t}-\alpha_k^{\dagger}  V_k(z)^* e^{i\nu_k t} \right].
\label{excitationb}
\end{eqnarray*}
Substituting Eqs.~(\ref{excitationa}) and (\ref{excitationb}) into
Eq.~(\ref{1dgpeq}), we obtain
\begin{eqnarray}
\left(-\frac{\hbar}{2m}\frac{\partial^2}{\partial z^2} + \eta
\right)U_k -\eta V_k &=& \nu_k U_k \nonumber \\
\left(-\frac{\hbar}{2m}\frac{\partial^2}{\partial z^2} + \eta
\right)V_k -\eta U_k &=& -\nu_k U_k \label{excitationeq}
\end{eqnarray}
where we have defined
\begin{equation}
\eta=g_{2,z} |u_0|^2 - 2 |g_{3,z}| |u_0|^4.
\end{equation}
Considering plane-wave excitations $U_k = (1/L)\exp(ikz)$ and
$V_k = (1/L)\exp(-ikz)$, Eq.~(\ref{excitationeq}) gives the
excitation frequencies
\begin{equation}
\nu_k=\sqrt{\frac{\hbar k^2}{2m}\left(\frac{\hbar k^2}{2m} + 2
\eta \right)},
\end{equation}
which reduce to the familiar Bogoliubov spectrum in the absence of
three-body excitations. In that limit, the excitation energies are
always positive and the condensate is stable. However, the
situation changes completely when $g_{2,z} < 2 |g_{3,z}| |u_0|^2$,
or $\eta < 0$, in which case the excitation energies can become
imaginary. Physically, such complex frequencies imply that
spatially modulated perturbations can grow with time, that is, a
modulational instability may occur. The maximum gain occurs for
the wave number $k$ such that $\hbar k^2/2m = |\eta|$, which
yields the spatial period of the most unstable wave as $L_m =
\sqrt{2 \pi^2 \hbar/(m|\eta|)}$. Provided that the length of the
BEC satisfies $L > L_m$, it will then become modulationally
unstable due to the attractive three-body interactions. As a
result, a spatial inhomogeneity will develop in the condensate
density profile along the $z$-axis. Three-body interactions impose
a fundamental limit on the linear densities that can be stably
propagated and manipulated in the waveguide.

Consider a $^{87}$Rb BEC as an example. The two-body interaction
coefficient of $^{87}$Rb is $g_2 = 4 \pi \hbar a_s/m \sim 4.95
\times 10^{-11}$ cm$^3$/s. The theoretical estimate of the
three-body interaction coefficient is $|g_3| \sim 10^{-26}
-10^{-27}$ cm$^6$/s. The onset of the modulational instability
occurs then at $|g_3| \rho_0\sim g_2/2 $, with $\rho_0$ being the
three-dimensional density of the condensate. This yields a
threshold density of the order of $10^{15} - 10^{16}$ cm$^{-3}$
beyond which the condensate becomes dynamically unstable. We
remark that such densities might have already been achieved in
recent experiments. The corresponding modulational length is on
the order of a few microns.

\section{Numerical simulations}
In this Section we present numerical simulations that illustrate
the effect of the modulational instability on the condensate
dynamics. In particular, we consider the case the magnitude of
the bias magnetic field $B_{\rm bias}$ is decreased to lower a
magnetically trapped gas towards the atom chip surface. We assume
that the initial trapped gas is stable against modulational
instabilities. As the condensate approaches the surface of the
chip, its linear density increases concomitantly with the
increasing transverse trapping frequency. The gas can then become
modulationally unstable. Our numerics give illustrative examples
of the development of this instability.

For convenience of numerical calculation, we introduce the
dimensionless variables $\tau=\omega_z t,\zeta=z/l_z,$ and
$\Phi(\zeta,\tau)=\sqrt{l_z}u$, to yield
\begin{eqnarray}
i \frac{\partial \Phi}{\partial \tau} &=& \left (-\frac{1}{2}
\frac{\partial^2 }{\partial \zeta^2}+
\frac{1}{2} \zeta^2 \right ) \Phi \nonumber \\
&+& \beta_2 f(\tau)|\Phi|^2 \Phi +\beta_3 f^2(\tau)|\Phi|^4 \Phi,
\label{1dgpnm}
\end{eqnarray}
where
\begin{equation}
\beta_2 = \frac{(g_2 N/2\pi l_z^3)}{\hbar \omega_z} \xi_0, \quad
\beta_3 = \frac{(g_3 N^2/3\pi^2 l_z^6)}{\hbar \omega_z} \xi_0^2 .
\end{equation}
Here we have written
$$
\xi(\tau)=\omega_r(\tau)/\omega_z=\xi_0 f(\tau),
$$
$\xi_0=\omega_r(0)/\omega_z$ being the initial ratio of the
transverse and longitudinal trapping frequencies. The function
$f(\tau)$ incorporates the time variation of the magnetic bias
field. We choose $\xi_0$ such that three-body interactions are
initially negligible for the trapped gas and the condensate is
initially stable. As it is lowered the transverse trapping
frequency and linear density increase, and so $f(\tau)$ increases
from unity. We note from Eq. (\ref{1dgpnm}) that the two-body
interaction term is proportional to $f(\tau)$ and the three-body
term is proportional to $f^2(\tau)$, so that as the BEC
approaches the surface the importance of three-body interactions
increases, as we have seen.

We take parameters appropriate to a $^{87}$Rb BEC composed of
$N=10^4$ atoms, $m=1.44\times 10^{-25}$ kg, $g_2 = 4.95 \times
10^{-11}$ cm$^3$/s, $g_3 = - 4 \times 10^{-26}$ cm$^6$/s, with a
longitudinal trap frequency $\omega_z=2\pi\times 14$ rad/s, giving
$l_z=2.8$ $\mu$m \cite{note}. With these parameters we find
$\beta_2=375$ and $|\beta_3|=270$. We choose the initial
transverse trapping frequency such that
$\xi_0=\omega_r(0)/\omega_z=10$, and assume a linear ramp of
$\omega_r(\tau)$, $f(\tau)=1+\alpha \tau$, with $\alpha=1.5$. The
initial macroscopic wave function $\Phi(\zeta,0)$ is chosen as the
ground state $\chi(\zeta)$ of the trap, neglecting three-body
interactions. An approximation to the ground state is obtained in
the Thomas-Fermi approximation by setting
$\Phi(\zeta,\tau)=\exp(-i\gamma \tau)\chi(\zeta)$ and imposing the
normalization condition $\int d\zeta |\chi(\zeta)|^2=1$ to yield
\begin{eqnarray}
|\chi(\zeta)|^2 &=& {1 \over \beta_2}\left(\mu-\zeta^2/2 \right)
\theta\left(\mu-\zeta^2/2 \right ), \nonumber \\ \gamma &=& {1
\over 2}\left ( {3\beta_2 \over 2} \right )^{2/3} ,
\end{eqnarray}
with $\theta(y)$ the Heaviside step function. Our condition that
the transverse trapping frequency varies slowly compared to the
longitudinal dynamics requires $\gamma>>\alpha$, which is
satisfied by our chosen parameters.

\begin{figure}
\includegraphics*[width=1.0\columnwidth,
height=0.5\columnwidth]{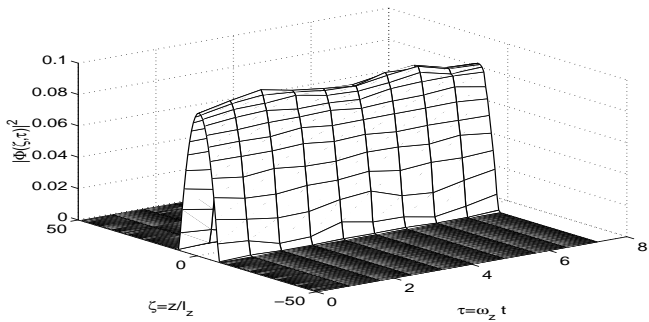} \caption{Scaled density
$|\Phi(\zeta,\tau)|^2$ versus $\zeta$ and $\tau$ for the case
with no ramping of the transverse trapping frequency
($\alpha=0$).} \label{fig1}
\end{figure}

Figure \ref{fig1} shows the evolution of the scaled linear density
$|\Phi(\zeta,\tau)|^2$ versus $\zeta$ and $\tau$ in the absence
of ramping of the transverse trapping frequency ($\alpha=0$). The
density profile remains intact in time, except for a small
modulation that arises from the fact that the initial macroscopic
wave function is the ground state without three-body
interactions, whereas the numerical simulation in Fig. \ref{fig1}
incorporates three-body interactions. This results in some small
rearrangement of the density profile. The robustness of the
density profile to the inclusion of the three-body interactions
shows that the initial state is stable and the three-body effects
are not important for these initial parameters. This also
justifies our use of this particular initial wave function.

\begin{figure}
\includegraphics*[width=1.0\columnwidth,
height=0.5\columnwidth]{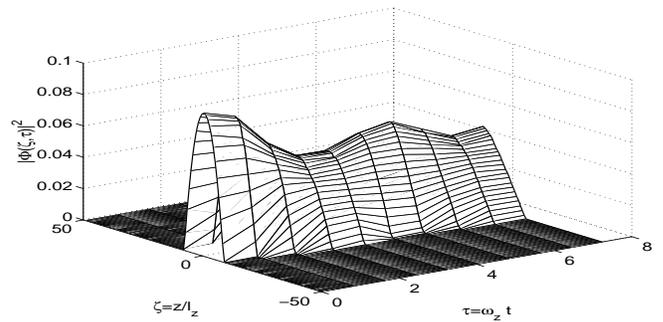} \caption{Scaled density
$|\Phi(\zeta,\tau)|^2$ versus $\zeta$ and $\tau$ for the case
with ramping of the transverse trapping frequency
($\alpha=1.5$).} \label{fig2}
\end{figure}

This result should be compared and contrasted with Fig.
\ref{fig2}, which shows the evolution of the scaled linear
density $|\Phi(\zeta,\tau)|^2$ versus $\tau$ for $\tau=0
\rightarrow 7.5$ for a situation where the transverse trapping
frequency is ramped ($\alpha=1.5$). As the transverse trapping
frequency $\xi(\tau)=\omega(\tau)/\omega_z=10\times(1+1.5\tau)$
increases in time, the initial effect is a broadening of
condensate width. This is a consequence of the fact that the
dominant repulsive two-body interactions are becoming stronger.
The density undulations seen in Fig. \ref{fig2} arise from the
interplay between the attractive harmonic trapping and the
increasing two-body repulsion.

\begin{figure}
\includegraphics*[width=1.0\columnwidth,
height=0.5\columnwidth]{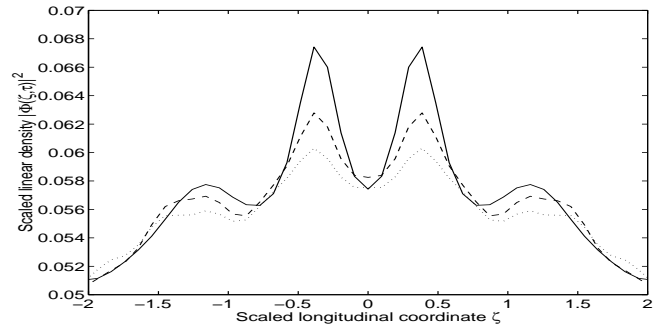} \caption{Scaled density
$|\Phi(\zeta,\tau)|^2$ versus $\zeta$ for times $\tau=$7.64
(dotted line), $\tau$=7.68 (dashed line), and $\tau$=7.72 (solid
line).} \label{fig3}
\end{figure}

As time increases further, the role of the attractive three-body
interactions starts to become important, with the onset of the
modulational instability kicking for $\tau>7.5$. Figure \ref{fig3}
shows the development of the instability around the center of the
trap at the dimensionless times $\tau=$7.64 (dotted line), $\tau$
=7.68 (dashed line), and $\tau$=7.72 (solid line). A spatial
modulation of the condensate density of period $L/l_z\simeq 0.8$
is clearly seen to be developing, resulting in a fragmentation of
the density profile of the trapped gas.

One can compare this period with the analytical predictions of
Sec. III by treating the center of the BEC as approximately
homogeneous. Converting to dimensionless units, the period of the
most unstable wave vector becomes
\begin{equation}
\frac{L_m}{l_z} \simeq \sqrt{ \frac{2\pi^2}{|\beta_2f_{\rm
max}n_c-2|\beta_3| f_{\rm max}^2n_c^2|} } , \label{Lm}
\end{equation}
where $f_{\rm max}=13$ is the value of $f(\tau=8)$ for times in
the vicinity of where the instability develops, and $n_c \simeq
0.06$ is the value of the scaled linear density at the trap
center. From this we obtain $L_m/l_z \approx 0.74$, in reasonable
agreement with the period $0.8$ between the peaks in Fig.
\ref{fig3}. We remark that the threshold for the modulational
instability occurs when the denominator in Eq. (\ref{Lm})
vanishes, that is, $|\beta_2f_{\rm max}n_c-2 |\beta_3| f_{\rm
max}^2n_c^2|=0$.

For times $\tau>8$ the main density peaks generated by the
modulational instability in Fig. \ref{fig3} continue growing, and
the numerical simulations quickly break down thereafter. Indeed,
the numerical solutions indicate that the density peaks are
undergoing a collapse whereby the density locally increases
towards infinity within a finite time. Abdullaev {\it et. al.}
\cite{AbdGamTom01} have previously shown that collapse is possible
in a trapped gas with repulsive two-body interactions. This
collapse is analogous to the corresponding case for two- or
three-dimensional condensates, where a collapse occurs for
condensates large enough that the attractive two-body interactions
overcome the ``quantum pressure'' associated with the trap ground
state. By contrast, the present system is effectively
one-dimensional, (in $\zeta$) but with a combination of repulsive
two-body interactions and attractive three-body interactions, and
it is the competition between these two mechanisms that leads to
condensate collapse for sufficiently high densities. This places a
fundamental limitation on the linear densities that can be stably
sustained in atom microtraps and waveguides.

In practice, the predicted collapse will be regularized by some
additional physical process, such as three-body losses as in the
case of higher-dimensional collapse in the Bose-Novae. We
conjecture that a form of soliton turbulence will ensue
\cite{Zak,Dya}. In addition, as the linear density continues to
grow the mean-field energy will eventually exceed the transverse
mode energy, in which case multiple transverse modes will be
excited thereby arresting the collapse.
\section{Collapse and modulational instability}
In this section, we present a discussion that reinforces the
argument that the modulational instability eventually leads to
condensate collapse. We proceed by reexamining the generalized
Gross-Pitaevskii Eq. (\ref{1dgpnm}), which can be interpreted as
resulting from the effective potential
\begin{eqnarray}
U_{eff}(\zeta)=\frac{1}{2}\zeta^2 + \beta_2 f |\Phi|^2 + \beta_3
f^2  |\Phi|^4,
\end{eqnarray}
Consider a normalized Gaussian approximate trial solution of width
$l$
\begin{eqnarray}
\Phi(\zeta) \approx \left(\frac{1}{\pi l^2}\right)^{1/4}\,
e^{-\zeta^2/(2l^2)}, \label{trial}
\end{eqnarray}
which relates the peak density $|\Phi(0)|^2$ at trap center to $l$
via
\begin{equation}
n_c=|\Phi(0)|^2 = 1/\sqrt{\pi l^2}.
\label{density-width}
\end{equation}
Near the center, and evaluating the potential for parameters
$f_{\rm max}$, $n_c$ in the vicinity of where the modulational
instability develops, we have
\begin{widetext}
\begin{eqnarray}
U_{eff}(\zeta)\approx\left(\beta_2 f_{\rm max} n_c - |\beta_3|
f^2_{\rm max} n_c^2 \right) + \frac{\zeta^2}{l^2}
\underbrace{\left(\frac{l^2}{2} - \beta_2 \pi f_{\rm max} n_c
 + 2 |\beta_3| \pi f_{\rm max}^2 n_c^2 \right)}+ ...
\end{eqnarray}
\end{widetext}
We are interested in the under-braced term, which controls the
dominant confinement properties due to the combined effects of the
linear trap and two- and three-body interactions. At densities
low enough for two-body interactions to dominate over three-body
processes, the two-body collisions oppose the focusing of the
trap, and their balance produces a stable and confined solution.
In contrast, when attractive three-body interactions dominate,
both the trap and three-body interactions work to compress the
density, causing a strong focusing of the density distribution
and collapse. We therefore anticipate that the cross-over from
stability to collapse will occur when the two- and three-body
focusing terms in the under-braced term cancel each other,
\begin{eqnarray}
\beta_2 \pi f_{\rm max} n_c
 - 2 |\beta_3|  \pi f_{\rm max}^2 n_c^2  = 0 .
\end{eqnarray}
This coincides precisely with the modulational instability
threshold from Eq. (\ref{Lm}). Thus, the appearance of a
modulational instability also signals that the condensate will
collapse if the trapped gas is too close to the atom chip surface.
This conclusion is also in agreement with the results of Abdullaev
{\it et. al.} \cite{AbdGamTom01} regarding the stability of
trapped gases. We remark that although we have considered a
magnetically trapped gas, the same conclusions clearly apply to an
optically trapped gas at sufficiently high densities.

We can also use a variational approach to treat the collapse of
the condensate more rigorously. Take Eq.~(\ref{trial}) as the
variational wave function with the width $l$ being the variational
parameter. The energy functional associated with the dimensionless
Schr\"{o}dinger Eq.~(\ref{1dgpnm}) is then given by
\begin{widetext}
\begin{eqnarray}
E(l) &=& \int dz \, \Phi(\zeta)\left[
-\frac{1}{2}\frac{\partial^2}{\partial \zeta^2} +\frac{1}{2}
\zeta^2 + \frac{1}{2}\beta_2 f |\Phi(\zeta)|^2 +\frac{1}{3}
\beta_3 f^2 |\Phi(\zeta)|^4
\right] \Phi(\zeta), \nonumber \\
&=& \frac{l^2}{4} + \frac{\cal A}{l}- \frac{\cal B}{l^2},
\label{EFunctional}
\end{eqnarray}
\end{widetext}
where we used Eq. (\ref{density-width}), ${\cal A} = \beta_2
f/(2\sqrt{2\pi})$ and ${\cal B} = |\beta_3| f^2/(3\sqrt{3}\pi)-
1/4.$

From Eq.~(\ref{EFunctional}), we have that $E(l)\rightarrow
-\infty$ as $l\rightarrow 0$ provided that ${\cal B}>0$. This
means that under such a condition, the condensate is not stable
against collapse. However the system can still become metastable
if $E(l)$ possesses a local minimum at finite $l$. This is
analogous to the metastability of an attractive condensate with a
sufficiently small number of atoms \cite{metastable}. The
condition for the existence of a local minimum is $dE(l)/dl=0$ and
$d^2 E(l)/d^2 l
>0$, which yields
\begin{eqnarray*}
F(l) &\equiv & \frac{1}{2} l^4 -{\cal A}l + 2{\cal B}=0 ,\\
l &>& \frac{8{\cal B}}{3{\cal A}}.
\end{eqnarray*}
It can be shown that the above conditions are satisfied if and
only if
\begin{equation}
F\left(\frac{8{\cal B}}{3{\cal A}} \right)= \frac{1}{2} \left(
\frac{8{\cal B}}{3{\cal A}} \right)^4 - \frac{2}{3} {\cal B} < 0.
\label{condition}
\end{equation}
The local minimum vanishes when the above inequality becomes an
equality. For the parameters of our numerical calculations this
happens at $\tau \approx 8.4$ which is in reasonable agreement
with the onset of collapse $\tau>8$ found in the simulations.
This not only further establishes the connection between the
modulational instability and condensate collapse, but also
provides the condition for the threshold of the instability
through Eq. (\ref{condition}).
\section{Summary and conclusions}
Generic microtraps and atomic waveguides based on chip technology
are characterized by an increased level of transverse confinement,
and hence increased atomic density, as their distance from the
chip is reduced. As a result, three-body collisions become
increasingly important. We have shown that for attractive
collisions they can lead to a modulational instability in the
dynamics of the condensate, and its eventual collapse, in
agreement with the prior work of Abdullaev {\it et. al.}
\cite{AbdGamTom01} on the stability of trapped gases. The physics
underlying this collapse, which involves the competition between
repulsive two-body collisions and attractive three-body
collisions, is reminiscent of the physics underlying the collapse
of condensates with attractive two-body interactions, except that
in that latter case, the competition is between the effect of
collisions ant the quantum pressure associated with the trap
ground state. In both cases, though, the competition leads to a
fundamental limit on the size of condensates that can be stably
trapped and coherently manipulated, since the modulational
instability and collapse correspond to a form of spatial
fragmentation which will persist for trapped gases even though
current technical sources of noise are eliminated.

In addition to predicting a fundamental limit in integrated atom
optics, the present study also opens up intriguing new directions
of investigation. For instance, we recall that the use of Feshbach
resonances to switch the sign of the scattering length from
positive to negative has lead to the discovery of fascinating
dynamical effects such as Bose-Novae in three-dimensional
condensates. Similar studies, but in one dimension, should now
become possible simply by modulating either the bias field $B_{\rm
bias}$ or the current $I$ in wire microtraps. In practice, the
predicted collapse, and the subsequent dynamics of the collapsing
filaments, will be regularized by some additional physical
process, such as three-body losses as in the case of
higher-dimensional collapse in the Bose-Novae, or excitation of
higher-order transverse modes as the mean-field energy rises due
to collapse. In future work we shall investigate the detailed
dynamics of this one-dimensional Bose-Novae beyond the initial
growth of the modulational instability considered here.
\acknowledgments This work is supported in part by the US Office
of Naval Research under Contract No. 14-91-J1205 and No.
N00014-99-1-0806, by the National Science Foundation under Grant
No. PHY98-01099, by the US Army Research Office, by the NASA
Microgravity Fundamental Physics Program and by the Joint Services
Optics Program.


\begin{references}
\bibitem{note} For numerical reasons, a small number
of atoms is used here so that the threshold of the modulational
instability occurs at a relatively small wave number and hence can
be spatially resolved over the width of the condensate.
\end{references}
\end{document}